\documentclass[amsmath,amssymb,reprint,aps,prl,showpacs,superscriptaddress,floatfix]{revtex4-1}

\usepackage{graphicx}
\usepackage{color}
\usepackage{epstopdf}
\usepackage{braket}
\usepackage{siunitx}
\usepackage[normalem]{ulem}

\begin{document}

\title{Magnetic anisotropy in Shiba bound states across a quantum phase transition}
\author{Nino Hatter}
\author{Benjamin W. Heinrich}
\author{Michael Ruby}
\affiliation{Fachbereich Physik, Freie Universit\"at Berlin,
                 Arnimallee 14, 14195 Berlin, Germany.}
\author{Jose I. Pascual}
\affiliation{Fachbereich Physik, Freie Universit\"at Berlin,
                 Arnimallee 14, 14195 Berlin, Germany.}
\affiliation{CIC nanoGUNE and Ikerbasque, Basque Foundation for Science,
						     Tolosa Hiribidea 78, Donostia-San Sebastian 20018, Spain.}
\author{Katharina J. Franke}
\affiliation{Fachbereich Physik, Freie Universit\"at Berlin,
                 Arnimallee 14, 14195 Berlin, Germany.}
\date{\today}

\begin{abstract}
The exchange coupling between magnetic adsorbates and a superconducting substrate leads to Shiba states inside the superconducting energy gap and a Kondo resonance outside the gap. The exchange coupling strength determines whether the quantum many-body ground state is a Kondo singlet or a singlet of the paired superconducting quasiparticles. Here, we use scanning tunneling spectroscopy to identify the different quantum ground states of Manganese phthalocyanine on Pb(111). We observe Shiba states, which are split into triplets by magnetocrystalline anisotropy. Their characteristic spectral weight yields an unambiguous proof of the nature of the quantum ground state.

\end{abstract}


\maketitle

Magnetic adsorbates on a superconductor create a magnetic scattering potential for the quasi-particles of the superconductor. A single spin gives rise to so-called Yu-Shiba-Rusinov (Shiba) states~\cite{yu65,shiba68,rusinov69}. Recently, it was argued that hybridization of the Shiba states can lead to Shiba bands with nontrivial topological character~\cite{pientkaPRB13,nadjpScience14, RoentynenPRL15}. This is essential for the formation of Majorana modes, which have been detected in ferromagnetic chains of Fe atoms on a Pb(110) surface~\cite{nadjpScience14}. If not only one but several Shiba states are present, the hybridization will lead to a more complex band structure. 
Different origins of multiple Shiba states are discussed theoretically. These include different angular momentum scattering contributions, individual $d$ orbitals acting as separate scattering potentials, or low-energy excitations due to magnetic anisotropy or vibrations~\cite{ginsberg79, kunz80, flatte97PRL, golez12, ZitkoPRB11, KimPRL15}.  
However, experimentally, the origin of multiple Shiba states is often difficult to determine~\cite{jiPRL08, Ruby15}. 

Concomitantly with the formation of a Shiba state, the exchange coupling between the magnetic adsorbate and the substrate also drives the formation of a singlet Kondo state
~\cite{matsuura77, ZitkoPRB11,Franke11, bauer13}. If the Kondo energy scale $k_{\mathrm{B}}T_{\mathrm{K}}$ is much larger than the superconducting pairing energy $\Delta$, the "Kondo screened" state is the ground state.
On the other hand, if $k_{\mathrm{B}}T_{\mathrm{K}}\ll\Delta$, an unscreened "free spin" state with $S>0$ is the ground state. A Shiba resonance as fingerprint of this magnetic interaction can be found in the quasiparticle excitation spectra within the superconducting energy gap if $k_{\mathrm{B}}T_{\mathrm{K}}\sim\Delta$~\cite{YazdaniScience97}. This resonance corresponds to a transient state, in which an electron is added/removed to/from the ground state. Thus, the electron occupation changes and the spin is altered by $\Delta S=\pm1/2$.
In the "Kondo screened" case, the Shiba resonance is found with a binding energy $E_\mathrm{b}$ below the Fermi level $E_\mathrm{F}$. The weaker the exchange coupling and Kondo screening, the closer is the Shiba state to the Fermi level. The crossing of the Shiba state through $E_\mathrm{F}$ marks the quantum phase transition from the "Kondo screened" $S=0$ to the "free spin" state on the superconductor, {\it i.e.}, $S>0$. This transition occurs at $k_{\mathrm{B}}T_{\mathrm{K}}\sim0.3\Delta$ 
and has been described theoretically~\cite{matsuura77, BalatskyReview06, bauer13} and experimentally~\cite{deaconPRL10, Franke11,leeNNano14}.

Spin-$1/2$ systems feature one pair of Shiba resonances at $\pm E_\mathrm{b}$~\cite{YazdaniScience97}. If the adsorbate carries a higher spin, multiple Shiba states may appear inside the gap as discussed theoretically by \v{Z}itko and co-workers~\cite{ZitkoPRB11}. They argue that such systems may involve multiple Kondo screening channels with different coupling strengths $J_k$ and, hence, multiple Shiba states. 
Furthermore, a mutual coupling of the spins may lead to a splitting of the peaks in the excitation spectra in the presence of magnetic anisotropy.  Here, we resolve triplets of Shiba states on single paramagnetic molecules. A multitude of different adsorption sites provides access to a large range of magnetic coupling strengths with the substrate. We detect the splitting of the Shiba states even throughout the quantum phase transition from a "Kondo screened" to a "free spin" ground state. The intensities of the Shiba resonances yield an unambiguous proof of the nature of the spin states and their splitting by magnetic anisotropy.
The basic understanding of the influence of magnetic anisotropy on many-body interactions is crucial for the design of quantum states with controlled properties. Furthermore, its knowledge may provide interesting approaches for creating and addressing Majorana states in proximity coupled magnetic nanostructures~\cite{KimPRL15}.

\section{Results}
\subsection{Detection of Shiba states}

The effects of magnetic anisotropy on Shiba states can be explored experimentally by bringing a metal-organic molecule into contact with a superconductor. The organic ligand is responsible for the splitting of the spin states with $S\geq1$ of the transition metal core~\cite{Gatteschi06, tsukahara09}.   
Manganese phthalocyanine (MnPc) has a spin $S=3/2$ in gas phase and retains a magnetic moment on metal surfaces~\cite{ji10,StrozeckaPRL,BodeNanoLett14}. In particular, its magnetic moment interacts with the superconductor Pb and shows single Shiba states when measured at 4.5~K~\cite{Franke11}.

\begin{figure}[t]
  \includegraphics[width=0.48\textwidth,clip]{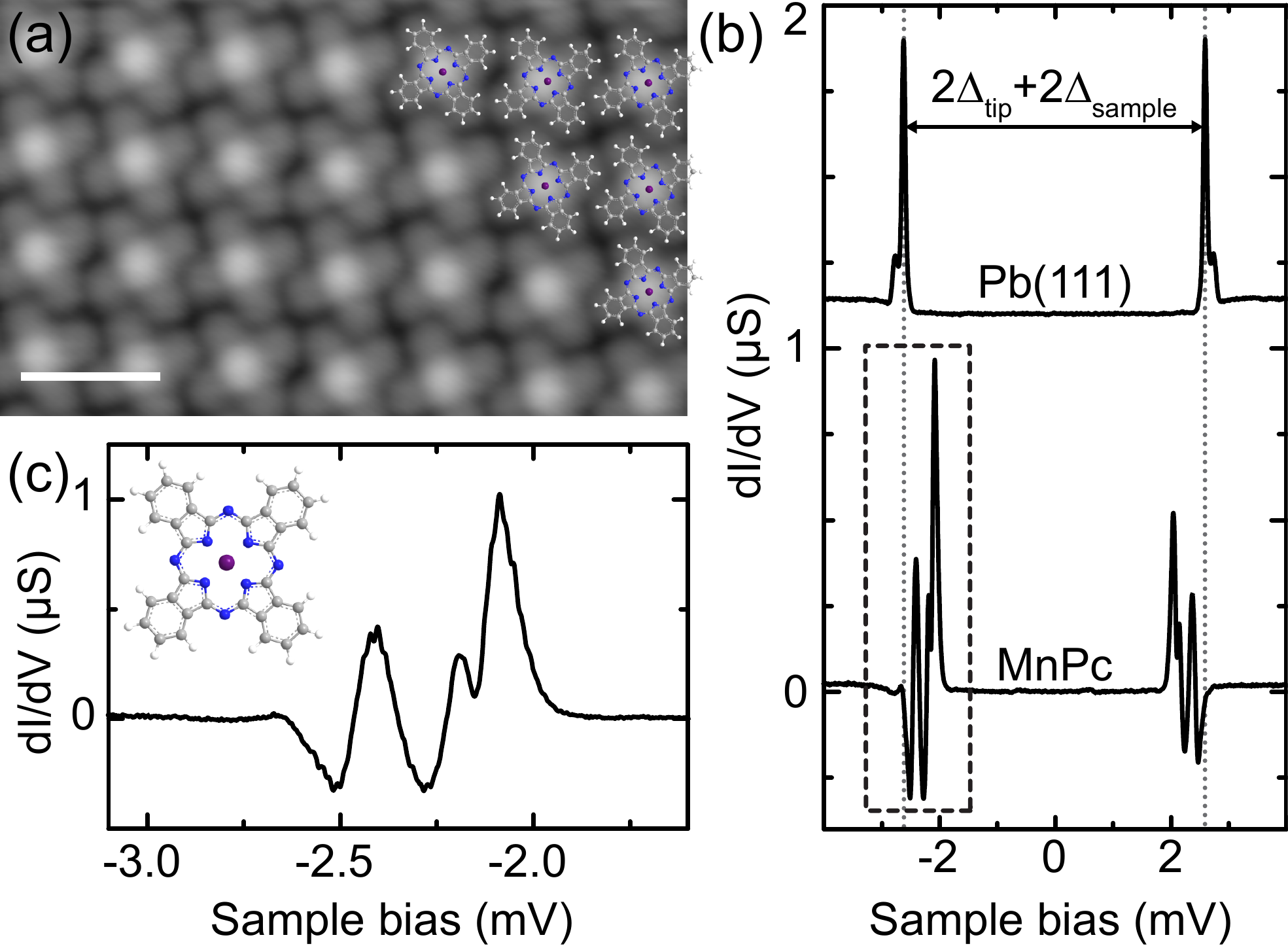}
  \caption{Manganese-phthalocyanine (MnPc) on Pb(111): (a) STM topography of a MnPc monolayer island on Pb(111) ($V=50~$mV, $I=200~$pA), scale bar is 2~nm.  (b) $dI/dV$ spectra on the pristine Pb(111) surface ({\it top}, offset for clarity) and on the center of a MnPc molecule ({\it bottom})  inside a molecular island (opening feedback loop at: $V=5~$mV, $I=200~$pA). 
(c) Zoom on the subgap excitations below $E_F$ in the MnPc spectrum in (b). The inset shows a MnPc structure model.} 
\label{fig1}
\end{figure}

Deposition of MnPc molecules at room temperature results in self-assembled, densely-packed monolayer islands [see Fig.~\ref{fig1}(a)]. The molecules appear clover-shaped with four lobes around a central protrusion of the Mn ion. The nearly square lattice of the molecular adlayer accommodates many different adsorption sites of the Mn core on the hexagonal Pb lattice. Consequently, this Moir\'e-like pattern involves variations in the electronic and magnetic coupling strength between adsorbate and substrate ~\cite{ji10,Franke11}. This rich system allows us to identify different quantum ground states with distinct fingerprints of their magnetic excitations.

\begin{figure*}[t]
  \includegraphics[width=0.96\textwidth,clip]{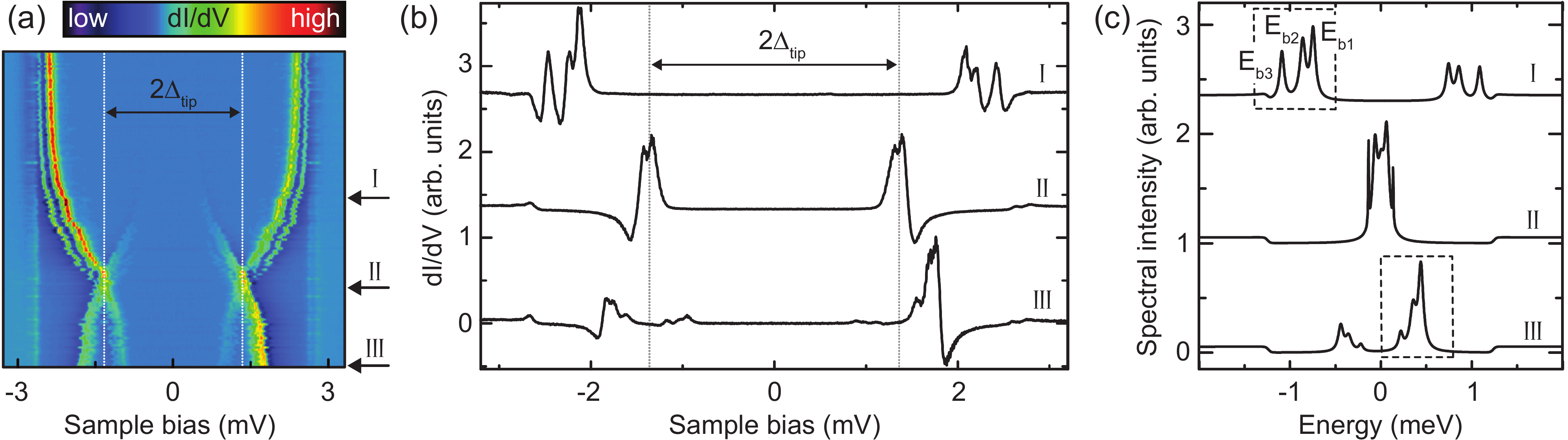}
  \caption{Differential conductance spectra of MnPc in Moir\'e-like pattern: (a) False color plot of $dI/dV$ spectra of $137$ MnPc molecules ordered by 
the energy of the most intense Shiba resonance (feedback: $V=5~$mV, $I=200~$pA). 
 (b) Spectra of three MnPc molecules with bound state energies in three different coupling regimes. 
 (c) Spectral intensity obtained by deconvolution of the spectra shown in (b). 
}
\label{fig2}
\end{figure*}

We use tunneling spectroscopy with a superconducting Pb tip at 1.2\,K to detect fingerprints of magnetic interaction of MnPc with the superconducting substrate. As a reference, we plot the differential conductance ($dI/dV$) spectrum of the bare Pb surface in Fig.~\ref{fig1}(b). A region of zero conductance around the Fermi energy ($E_\mathrm{F}$), {\it i.e.}, the superconducting gap, is framed by quasiparticle resonances at $eV=\pm (\Delta_\mathrm{sample} + \Delta_\mathrm{tip})=\pm$~2.63\,meV. The doubling of the size of the SC gap is due to the superconductor-superconductor tunneling geometry. 
The observed presence of the two quasiparticle resonances at each side of the gap is explained by the two-band superconductivity of the Pb single crystal as described recently~\cite{ruby14}.

Interestingly, the spectra on the MnPc molecules show two triplet sets of peaks inside the superconducting energy gap, which are symmetric in energy around $E_\mathrm{F}$, but asymmetric in intensity [Fig.~\ref{fig1}(b)]. 
In the limit of small tunneling rates -- as in our experiment -- the asymmetric intensity is an expression of the different hole and electron components of the Shiba wavefunctions~\cite{Ruby15}. The different weights arise from the particle-hole asymmetry in the normal state and an on-site Coulomb potential at the scattering site~\cite{flatte97PRL, salkola97prb,bauer13}.

The triplets consist of very sharp peaks ($50$ to $100~\mu$eV full width at half maximum), which are separated by up to $400~\mu$eV [{\it e.g.}, Fig.~\ref{fig1}(c)]. To observe such narrow peaks at $1.2$~K, a superconducting tip is required, because then the resolution is not limited by the Fermi-Dirac distribution anymore~\cite{ruby14, note1}.

The varying coupling strength within the Moir\'e-like structure leads to different bound state energies~\cite{Franke11}. We use this property to further investigate the origin of the splitting of the Shiba resonances and perform tunneling spectroscopy on more than 130 molecules.
All spectra exhibit two triplets of peaks, one in the bias voltage window $-2\Delta/e<V_\mathrm{bias}<-\Delta/e$ and one in $\Delta/e<V_\mathrm{bias}<2\Delta/e$. The additional resonances in the energy interval $[-\Delta_\mathrm{tip}, \Delta_\mathrm{tip}]$ are due to tunneling into/out of thermally excited Shiba states (Supplementary Fig.~1, Supplementary Note~1).
In Fig.~\ref{fig2}(a), we ordered the spectra according to the energy of the most intense Shiba resonance. The false color plot shows a collective "shift" of the Shiba triplets through the superconducting gap. The spectra can be categorized into three different regimes. For each of these, we plot a spectrum in Fig.~\ref{fig2}(b). In spectrum I, the intensity of the triplet is larger for tunneling out of the occupied states; in spectrum II, the peaks are close to $E_\mathrm{F}$, which hinders a clear distinction of the triplet; spectrum III exhibits larger intensity of the triplet when tunneling into unoccupied states. The energetic position of the higher intensity subgap peaks corresponds to the binding energy $E_\mathrm{b}$ of the Shiba states~\cite{matsuura77,Franke11,bauer13}. The order of the spectra from top to bottom thus represents a decreasing coupling strength $J$ with the substrate, which comes along with different adsorption sites. The three spectra are representative for the "Kondo screened" case [$E_\mathrm{b}<0$, spectrum I in Fig.~\ref{fig2}(b)], the "free spin" case ($E_\mathrm{b}>0$, spectrum III), and a case close to the quantum phase transition ($E_\mathrm{b}\approx0$, spectrum II). 

\subsection{Possible origins of a Shiba state splitting}
The collective shift of the Shiba states through a wide range of the gap (and even the quantum phase transition) suggests a correlated origin of the peaks within the triplet. In principle, three different scenarios may 
account for the occurrence of multiple Shiba bound states in a type I superconductor:
(i) different angular momentum scattering channels~\cite{ginsberg79,kunz80,flatte97PRL, jiPRL08, KimPRL15}, 
(ii) independent scattering at spins in different $d$ orbitals~\cite{ZitkoPRB11}, and (iii) bound state excitations coupled to other low-energy excitations, such as spin excitations or vibrations~\cite{ZitkoPRB11,golez12}.

Bound states, which originate from higher angular momentum scattering channels ($l=1,2, etc.$), always reside close to the gap edge, while the $l=0$ channel may shift through the superconducting energy gap depending on the coupling strength $J$~\cite{flatte97PRL, KimPRL15}. This is in contrast to our observation, and, hence, we can rule out (i) as possible origin for the split bound states.
In the case of independent scattering of spins in different $d$ orbitals (ii), a similar shift of all $E_{\mathrm{b}i}$ is unexpected, because $d$ orbitals exhibit different symmetries and interactions with the surface. 

For the coupling to other degrees of freedom at a similar energy scale (iii), a collective shift is expected. 
As we will show in the following, a detailed analysis of the intensities of the Shiba states allows us to unambiguously identify a magnetocrystalline origin of the splitting of the Shiba states as predicted by \v{Z}itko and co-workers~\cite{ZitkoPRB11} for $S\geq1$ systems.

\subsection{Shiba intensities as fingerprint of a quantum phase transition}
The evaluation of the Shiba intensities also sustains the assigned regimes of "Kondo screened" and "free spin" ground states. Such an analysis requires the spectral density of the molecule-substrate system, which can be directly related to the relative weight in the tunneling processes. For this, we remove the effect of the superconducting density of states of the tip by numerical deconvolution of the spectra as described in the Supplementary Note~2 (see Supplementary Fig.~2 for fit quality) [Fig.~\ref{fig2}(c)]. 

\begin{figure}[h]
  \includegraphics[width=0.48\textwidth,clip]{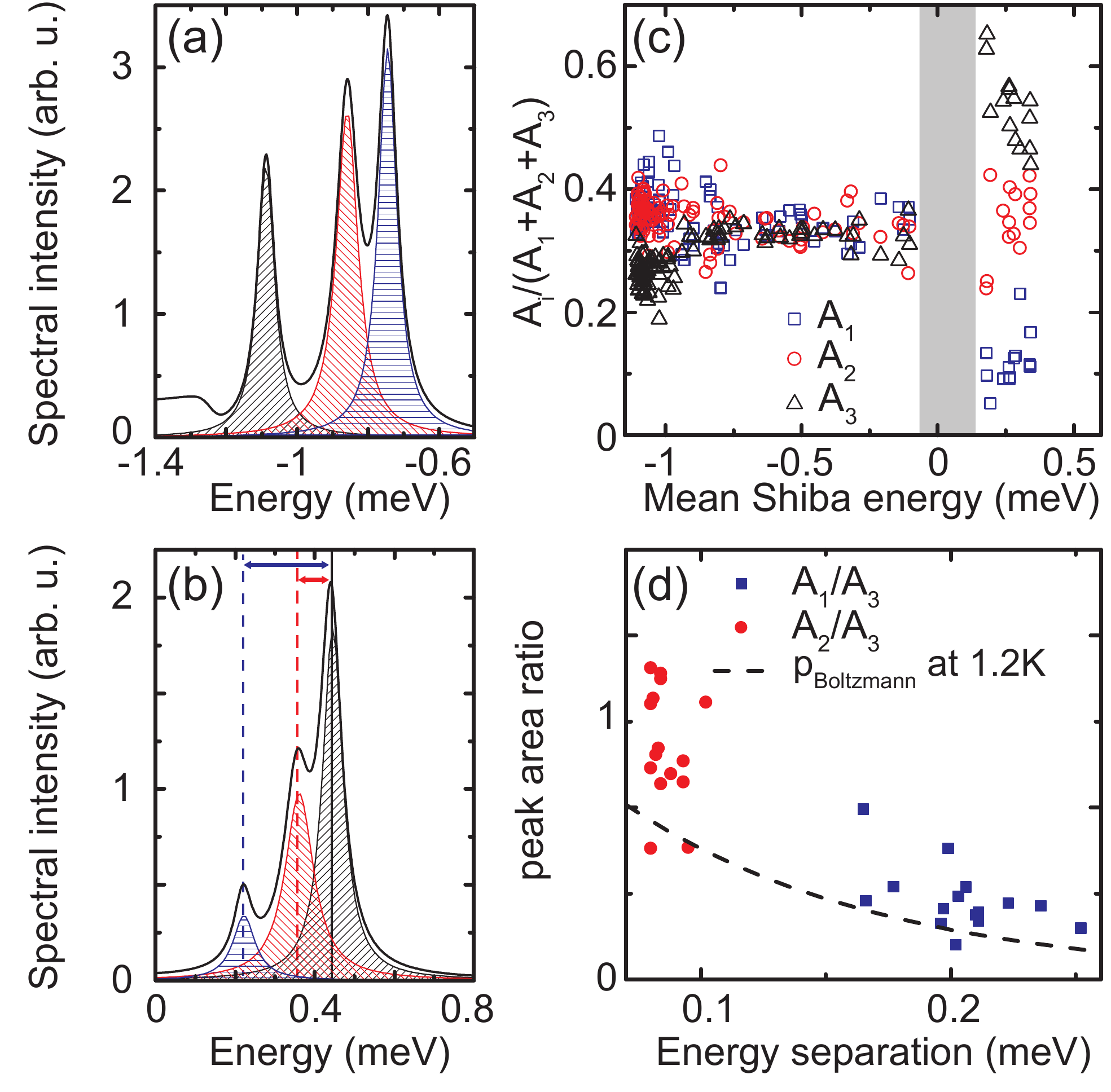}
  \caption{Shiba state analysis: (a) Zoom on the framed part of the deconvoluted spectrum I in Fig.~\ref{fig2}(c).
We model the spectrum with three Lorentzian peaks of different width (shaded in blue, red, and gray) and a broadened step function at the gap edge (see Supplementary Material for details). (b) As (a), but on spectrum III in Fig.~\ref{fig2}(c). (c) Peak areas relative to the sum of all three subgap peaks ($A_i/\sum_{i=1}^3{A_i}$) plotted vs. the mean energy ($E_{\mathrm{b}i}/\sum_{i=1}^3E_{\mathrm{b}i}$) of the respective set of subgap peaks. (d) Peak area ratios plotted vs. their respective energy splitting from all spectra in the "free spin" state. A Boltzmann distribution for $T=1.2$~K is sketched as dashed line.
}
\label{fig3}
\end{figure}

We observe a distinct change in the relative peak areas within the triplets when crossing the quantum phase transition. In the "Kondo screened" regime, {\it i.e.}, for Shiba states with negative binding energies, the individual peaks within a triplet exhibit equal areas [Fig.~\ref{fig3}(a)]. Thus, their relative areas $A_i/\Sigma_{j=1}^3A_j$ are close to $1/3$ [left part in Fig.~\ref{fig3}(c)].
 In contrast, in the "free spin" regime, {\it i.e.}, positive Shiba binding energies, the areas are considerably different [Fig.~\ref{fig3}(b) and (c)]. Here, the relative area of the subgap excitations decreases from the outermost, $E_{\mathrm{b}3}$, to the innermost, $E_\mathrm{{b}1}$~\cite{note2}. The ratio of peak areas decreases with increasing energy separation [Fig.~\ref{fig3}(d)]. Such a behavior is reminiscent of a Boltzmann distribution, indicating a thermal occupation of a split many-body ground state.

\begin{figure}[h]
  \includegraphics[width=0.48\textwidth,clip]{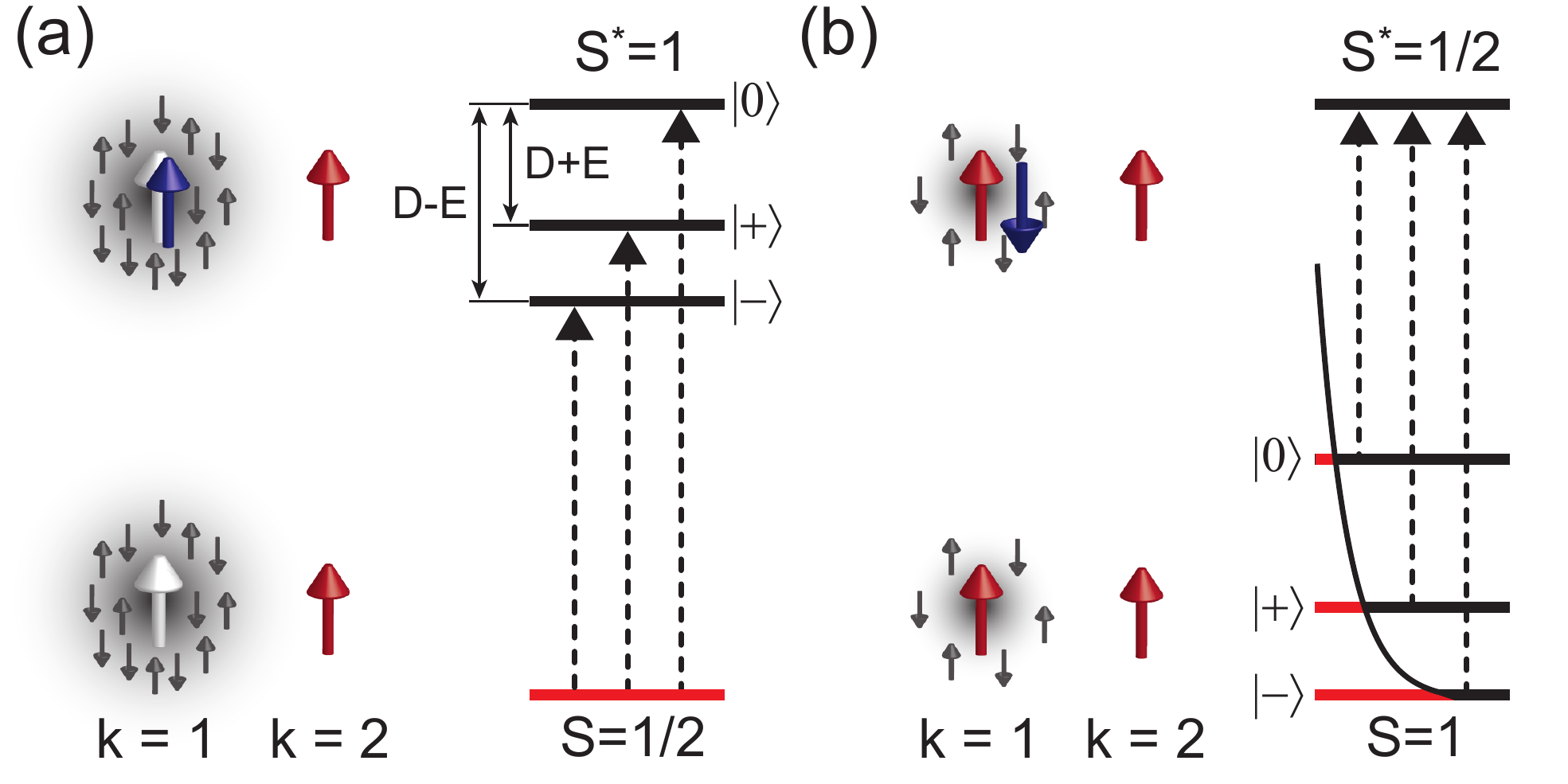}
  \caption{Schematic representation of the many-body ground and excited states and the corresponding energy level diagrams. (a) In the "Kondo screened" ground state the spin ({\it white}) in scattering channel $k=1$ is screened and the tunneling electron can enter with its spin ({\it blue}) parallel to the spin in $k=2$, increasing the excited state's total spin to $S^*=1$. The excitation scheme including the anisotropy-split excited state is shown on the right. (b) In the "free spin" ground state, the spin in $k=1$ ({\it red}) is only partially screened. The tunneling electron has to enter this state in an anti-parallel alignment, obeying the Pauli exclusion principle and reducing the spin to $S^* =1/2$. Red lines in the excitation scheme symbolize the thermal occupation of the anisotropy-split ground state.}
\label{fig4}
\end{figure}

The characteristic change in the relative peak areas at the point of the quantum phase transition is a direct fingerprint of the origin of the splitting. Independent bound states of different $d$ orbitals (ii) would not change their relative weight at the phase transition. Furthermore, the phase transition should not occur simultaneously for all scattering channels. Additional vibronic resonances (iii) would appear as satellite peaks at higher absolute energy for both ground states. Their spectral weight should scale with the electron-phonon coupling strength, albeit being substantially lower than the weight of the main resonance~\cite{golez12}. This is clearly not the case in our data. Hence, both scenarios can safely be ruled out.

\subsection{Assignment of scattering channels and anisotropy splitting of Shiba states}
To conclude on the correct model for the description of the multiple subgap resonances, we summarize the essential properties of the Shiba states: On the one hand, the equal peak areas for  the "Kondo screened" ground state reflect that our system is characterized by a single level  with three possible excitation levels of equal probability. On the other hand, the Boltzmann-like distribution of the areas in the "free spin" case indicate a triplet-split ground state with one excited state. We can correlate these levels to the magnetic interaction channels of the MnPc molecule with the substrate. The MnPc molecule carries a spin $S=3/2$ in gas phase~\cite{liao-inorg-chem-2005}. Theory predicts that on Pb(111), the spin in the $d_\mathrm{xz, yz}$ forms a singlet with the organic ligand states, which reduces the effective spin {\it seen} by the substrate's quasiparticles to $S=1$~\cite{jacob13}. The unpaired spin in the $d_\mathrm{z^2}$ orbital is subject to strong coupling with the electronic states of the substrate leading to sizable Kondo screening, whereas the spin in the $d_\mathrm{xy}$ orbital is not expected to show a significant coupling with the substrate~\cite{jacob13,BodeNanoLett14, KuegelPRB15}. The occurrence of the Shiba states in tunneling spectra is thus linked to the interaction of the $d_\mathrm{z^2}$ orbital with the substrate. 
We label this scattering channel as $k=1$ (sketches in Fig.~\ref{fig4}). 
The unscreened spin in the $d_\mathrm{xy}$ orbital (which we label as $k=2$) does not give rise to an observable Shiba state in agreement with the theoretical predictions~\cite{jacob13}. However, it couples to the spin in the $d_\mathrm{z^2}$ orbital and leads to an anisotropy splitting of the Shiba state with $k = 1$ as discussed below.

We can describe the whole set of spectra on the different molecules by these spin states and their interactions: The coupling strength $J_1$ of channel $k=1$ depends on the adsorption site of MnPc on the Pb(111) surface. In the case of strong coupling, the spin is totally Kondo screened [white arrow, $E_\mathrm{b} <0$, Fig.~\ref{fig4}(a)], but $k=2$ remains unscreened (red arrow). Hence, the effective total spin is reduced to $S=1/2$ by Kondo screening. Tunneling into the Shiba state reflects the excitation to $S^*=1$. 
In the case of weak coupling in $k=1$ [left red arrow, $E_\mathrm{b} >0$, Fig.~\ref{fig4}(b)], the total spin in  the ground state multiplet is $S=1$. The excited state probed by the Shiba resonance is a $S^*=1/2$ state, because the electron attached to $k=1$ must obey Pauli's exclusion principle and align anti-parallel.

Both spin-1 states, {\it i.e.}, $S^*=1$ and  $S=1$, can be split by magnetic anisotropy.
A breaking of spherical symmetry of the Mn orbitals by the organic ligand and the adsorption on a substrate leads to a splitting of these spin states~\cite{Gatteschi06,tsukahara09}. The corresponding Spin Hamiltonian $H_S=D S_{\mathrm{z}}^2 + E \left(S_{\mathrm{x}}^2 - S_{\mathrm{y}}^2\right)$,
where the $S_i$ are the spin operators in Cartesian coordinates, accounts for the axial anisotropy and additional rhombicity with the parameters $D$ and $E$, respectively. 
This yields a new set of spin eigenstates $\ket{0}$, $\ket{+}$, and $\ket{-}$, with the latter two being linear combinations of the former eigenstates with $m_\mathrm{s}=1$ and $m_\mathrm{s}=-1$~\cite{tsukahara09}.
 Schemes of the excitations observed in the tunneling spectra are shown in Fig.~\ref{fig4}(a,b). In the strongly coupled regime ("Kondo screened"), the excitation of the system yields three excited states with the energy splittings being related to the anisotropy parameters $D$ and $E$. It should be noted, that their values are not a  direct measure of the anisotropy energies of the molecule on the surface, but rather represent a renormalized value due to the many-body interactions with the substrate~\cite{ZitkoPRB11}. Since the energy separation between $E_{\mathrm{b}1}$ and $E_{\mathrm{b}2}$ is smaller than between $E_{\mathrm{b}2}$ and $E_{\mathrm{b}3}$, the anisotropy parameter $D$ is negative, which means easy-axis anisotropy.
For the weakly coupled system, {\it i.e.}, the "free spin" case, the ground state is split by anisotropy. The excitation spectra reflect transitions from the states $\ket{-}$, $\ket{+}$ and $\ket{0}$ into the excited $S^*=1/2$ state. 

The characteristic variations of the peak areas are also well captured in this scenario of anisotropy-split Shiba states: All three spin excitations in the "Kondo screened" regime are equally probable, therefore leading to the same relative peak area [Fig.~\ref{fig3}(c)]. On the other hand, a splitting of the ground state, as it is found in the "free spin" regime, leads to a Boltzmann occupation of the levels. 
In the zero temperature limit, which is discussed in Ref.~\cite{ZitkoPRB11}, only one excitation would be detected in the "free spin" regime. At finite temperature, the levels are occupied according to Boltzmann statistics. The excitation probabilities are proportional to the state occupation and should thus directly reflect this distribution. 
Our data in Fig.~\ref{fig3}(d) is in agreement with a Boltzmann distribution at 1.2\,K or slightly higher, which reflects a temperature-induced population of the split ground state.

\section{Discussion}
Our study shows the importance of anisotropy effects on the subgap excitations which determine the electron transport properties. They provide an unambiguous fingerprint of the nature of the ground and excited state throughout the whole range of magnetic interaction strengths, which drive the phase transition between the two quantum ground states. Although the anisotropy energies are renormalized by the coupling to the substrates quasiparticles, this method can be used to extract knowledge about magnetic anisotropy and, hence, about spin-orbit coupling of magnetic adsorbates on superconductors.

Peculiar consequences of the split Shiba states may occur for coupled magnetic impurities, which lead to the formation of extended Shiba bands. If the exchange coupling strength is similar to or smaller than the anisotropy energy, the Shiba bands are expected to reflect the splitting of the individual Shiba states.
Recently, subgap bands have gained particular importance in the search of topological phases and Majorana states in ferromagnetic chains coupled to an s-wave superconductor~\cite{nadjpScience14}. If an odd number of spin-polarized bands crosses the Fermi level, Majorana end states can form in the presence of an induced (non-trivial) topological gap. Considering that the energy level splitting amounts to about one third of the superconducting gap, one may expect that a split Shiba band structure affects the number of crossing bands and the topological gap width, which is also in the order of 100~$\mu$eV \cite{Rubyarxiv15}.

\section{Methods}
The experiments were carried out in a commercial {\sc Specs} JT-STM operating at a base temperature of 1.2~K and a base pressure below $10^{-10}$~mbar. The Pb(111) single crystal surface was cleaned by repeated cycles of Ne$^+$ sputtering and annealing to 430~K until a clean, atomically flat, and SC surface was obtained. From a Knudsen cell held at 673~K, MnPc was thermally evaporated onto the clean surface kept at room temperature, which then was directly transferred into the precooled STM. 

To gain energy resolution beyond the Fermi-Dirac limit of a normal metal tip, we indented the chemically-etched W tip into the clean SC surface applying $100$~V tip bias until a Pb covered, superconducting tip was obtained (energy resolution better than $45~\mu eV$)~\cite{ruby14}. 
The resulting spectrum as acquired on the clean Pb(111) surface is shown in Fig.~\ref{fig1}(b) {\it top}. 
We acquired $dI/dV$ spectra as a function of sample bias  under open-feedback conditions using conventional {\it lock-in} technique with a bias modulation of 15~$\mu V_{rms}$ at an oscillation frequency of 912~Hz.

\section{Acknowledgments}
We thank F. von Oppen for fruitful discussions. We gratefully acknowledge funding by the Deutsche Forschungsgemeinschaft through grant FR2726/4 and by the European Research Council through grant "NanoSpin".




\begin{thebibliography}{unsrt}
\bibitem{note1} Yet, a finite temperature leads to finite lifetime effects, which broaden the superconducting coherence peaks and subgap states. Hence, already at 4.5\,K, the triplet peaks overlap and give rise to a single peak~\cite{Franke11}.
\end{thebibliography}
\end{document}